\documentclass[aps,prl,reprint]{revtex4-1}

\usepackage{amsmath,amssymb,amsfonts,dcolumn,color,graphicx,graphics,latexsym,placeins,epsfig}

\newcommand{\be}{\begin{eqnarray}}
\newcommand{\ee}{\end{eqnarray}}
\newcommand{\ba}{\begin{eqnarray}}
\newcommand{\ea}{\end{eqnarray}}

\def\ba{\begin{eqnarray}}
\def\ea{\end{eqnarray}}
\def\be{\begin{equation}}
\def\ee{\end{equation}}

\def\nn{\nonumber}

\def\eps{\epsilon}

\newcommand{\calR}{{\cal R}}
\newcommand{\calH}{{\cal H}}
\newcommand{\lapl}{\mathop\Delta^{(3)}}

\def\R{{\mathcal R}}

\def\P{{\mathcal P}}

\begin{document}

\begin{flushright}
		YITP-15-115
	\end{flushright}
	
\title{Adiabaticity and gravity theory independent conservation laws for cosmological perturbations}
\author{ 
Antonio Enea Romano$^{a,b}$, Sander Mooij$^{c}$, and
Misao Sasaki$^{d}$}

\affiliation{
$^{a}$Instituto de F\'isica, Universidad de Antioquia, A.A.1226, Medellin, Colombia\\
$^{b}$Department of Physics, University of Crete,
71003 Heraklion,Greece,\\
$^{c}$Grupo de Cosmolog\'ia y Astrof\'isica Te\'orica,
Departamento de F\'{i}sica, FCFM, Universidad de Chile,
Blanco Encalada 2008, Santiago, Chile\\
$^{d}$Yukawa Institute for Theoretical Physics, Kyoto University, Kyoto 606-8502, Japan}

\date{December 18, 2015}

\begin{abstract}
We carefully study the implications of adiabaticity for the behavior of
cosmological perturbations. There are essentially three similar but
different definitions of non-adiabaticity: one is appropriate for
a thermodynamic fluid $\delta P_{nad}$, another is for a general matter field
$\delta P_{c,nad}$, and the last one is valid only on superhorizon scales.
The first two definitions coincide if $c_s^2=c_w^2$ where
$c_s$ is the propagation speed of the perturbation, while 
$c_w^2=\dot P/\dot\rho$. 
Assuming the adiabaticity in the general sense, $\delta P_{c,nad}=0$, 
we derive a relation between the lapse function in the comoving sli\-cing $A_c$
and $\delta P_{nad}$ valid for arbitrary matter field in any theory of gravity,
by using only momentum conservation.
The relation implies that as long as $c_s\neq c_w$,
the uniform density, comoving and the proper-time slicings
coincide approximately for any gravity theory and for any matter field
if $\delta P_{nad}=0$ approximately. 
In the case of general relativity this gives the equivalence
between the comoving curvature perturbation $\R_c$ 
and the uniform density curvature perturbation $\zeta$
on superhorizon scales, and their conservation.
This is realized on superhorizon scales in standard slow-roll inflation.

We then consider an example in which $c_w=c_s$, where $\delta P_{nad}=\delta P_{c,nad}=0$ 
exactly, but the equivalence between $\R_c$ and $\zeta$ no longer holds.
Namely we consider the so-called ultra slow-roll inflation. 
In this case both $\R_c$ and $\zeta$ are not conserved.
In particular, as for $\zeta$, we find that it is crucial to take into
account the next-to-leading order term in $\zeta$'s spatial gradient expansion
to show its non-conservation, even on superhorizon scales.
This is an example of the fact that adiabaticity (in the thermodynamic sense) is not always enough to ensure 
the conservation of $\R_c$ or $\zeta$.
\end{abstract}

\pacs{04.70.-s, 04.20.Ha, 04.50.Kd}
\keywords{Scalar-tensor theory, cosmological constant, cosmological event horizon, no hair}

\maketitle

\section{Introduction}
It is a well-known fact that in single-field slow-roll 
inflation \cite{infl2,infl3,infl4}, the comoving curvature perturbation 
$\R_c$ and the uniform density curvature perturbation $\zeta$ coincide and are
 conserved. In the seminal works \cite{Malik3,lms}, it was shown that 
requiring just energy conservation is enough to show the 
superhorizon conservation of $\zeta$ given that the non-adiabatic pressure 
$\delta P_{nad}$ vanishes, under the assumption that gradient terms are negligible.
Moreover, it was shown in \cite{Malik3} that for adiabatic perturbations, on 
superhorizon scales the comoving sli\-cing coincides with the uniform density slicing, 
as long as $\partial V / \partial \phi \neq 0$. As a result, 
$\zeta$ and $\R_c$ coincide and both are conserved on superhorizon scales.

Nevertheless, there are cases in which the conservation of $\zeta$ or $\R_c$ 
does not hold even for adiabatic perturbations. This seems to contradict the results quoted in the above.
 In this paper, we carefully study the meaning of
adiabaticity and clarify how these seemingly contradictory statements
are reconciled. For this purpose, we first introduce three different
definitions of adiabaticity. 
Then we study the energy-momentum conservation laws for arbitrary matter 
and derive several useful relations among gauge-invariant variables, 
independent of the theory of gravity. 
We find a few useful formulas that relate some of the gauge-invariant variables
to each other. Then we specialize to the case of general relativity
and discuss the meaning of the conservation of $\zeta$ and $\R_c$ in detail.
Finally we study so-called ultra slow-roll inflation as an interesting non-trivial example in which the
superhorizon conservation of  $\zeta$ or $\R_c$ does not hold even for
an exactly adiabatic perturbation, $\delta P_{nad}=\delta P_{c,nad}=0$.

Throughout this paper the dot denotes the proper-time derivative ($\dot{}=d/dt$)
and the prime the conformal-time derivative ($\prime{}=d/d\eta$),
where $dt=a d\eta$, and the proper-time and
conformal-time Hubble expansion rates are respectively
denoted by $H=\dot a/a$ and $\calH=a'/a=\dot a$. 
\section{Adiabaticity: several definitions}
\label{sec:defs}
Let us consider several definitions of (non)-adiabaticity. 
Adiabaticity is apparently a term from thermodynamics.
Therefore ori\-gi\-nal\-ly it is meaningful only when
the basic matter variables such as the energy density and pressure
are thermodynamic.  As can be seen from the perturbed energy and 
momentum conservation equations for a perfect fluid with equation of 
state $P=P(\rho)$,  adiabatic perturbations move with the speed of sound $c_w$, 
given by
\be
c_w^2 \equiv \frac{P'}{\rho'}.
\ee
For a perfect adiabatic fluid, we therefore have $\delta P = c_w^2 \delta \rho$. 
Then it seems natural to define the non-adiabatic pressure as 
\be
\delta P_{nad}\equiv \delta P - c_w^2 \delta\rho, \label{dpnadther}
\ee
which is gauge invariant and vanishes for a perfect fluid. This is the definition
 used in \cite{Malik3,lms}, and in much of the literature.

However, the early universe is for sure not in thermal equilibrium, so one can 
question the above definition based on thermodynamics. In fact, when the universe 
is dominated by a scalar field, it makes more sense to talk about the propagation 
speed $c_s$ of that scalar field (the phase speed of sound, see 
also \cite{Christopherson:2008ry}), defined on comoving slices via
\be
c_s^2 \equiv \left(\frac{\delta P}{\delta \rho}\right)_c. \label{cssq}
\ee
One is then led to define the non-adiabatic pressure as
\be
\delta P_{c,nad}\equiv\delta P_c-c_s^2\delta\rho_c. \label{dpnadstr}
\ee

For a fluid, one has $c_s=c_w$ and both definitions coincide. However, this is in general not true. For a minimally coupled scalar field one has, for example, 
\be
c_w^2 = -1+\frac{2\eps}{3}-\frac{\eta}{3} , \qquad c_s^2=1,
\ee
with $\eps,\eta$ the usual slow-roll parameters. In this sense, 
the second definition is more general: It can apply
both to a fluid and to a scalar field, hence should be regarded
as the proper definition of adiabaticity. Therefore
we focus on the perturbation which satisfies $\delta P_{c,nad}=0$
in this paper. As a consequence, for the first definition we then have (in agreement with \cite{Christopherson:2012kw})
\be
\delta P_{nad} = (c_s^2-c_w^2) \delta \rho_c. \label{dpnad}
\ee


The third definition which is commonly used in the inflationary
cosmology is about the stage when the so-called growing mode
of the perturbation dominates. As discussed in the above, 
the adiabatic perturbation would generally satisfy a second-order
differential equation. Hence when it is Fourier decomposed
with respect to the spatial comoving wavenumber $k$,
there will be two independent solutions for each $k$-mode.
Usually what happens is that as the mode goes out of the Hubble horizon
during inflation,
one of the solutions (the decaying mode) dies out,
 and the other mode (the growing mode) dominates.
It turns out that this growing mode approaches a constant in
the superhorizon limit when expressed in terms of the curvature
 perturbation on comoving slices $\calR_c$ (or equally of the one
 on uniform energy density slices $\zeta$).
When the universe enters this stage where the growing mode
dominates, the evolution of the universe thereafter is unique. 
In other words, if we denote the time after which the universe
is in this growing mode dominated stage by $t_a$, given the state
of the universe at some later but arbitrary time $t_b$ ($>t_a$), one
can always recover the initial condition at $t=t_a$ uniquely because
the decaying mode is completely negligible during the whole stage
of evolution.
It is said that when this is the case the universe has arrived at the 
adiabatic stage (or the adiabatic limit). In particular, when the
universe is dominated by a scalar field whose evolution is well
described by the slow-roll approximation, this stage is reached
as soon as the scale of the perturbation leaves out of the horizon.

The above, third definition is different from the previous two
definitions in that it applies only to the stage when the wavelength
of the perturbation is much greater than the Hubble horizon.
Nevertheless, as long as we are interested in superhorizon scale
perturbations, the adiabaticity conditions for both of the previous two
cases will be approximately satisfied if the universe is in the adiabatic
limit. Namely, both $\delta P_{nad}$ and $\delta P_{c,nad}$ will 
be of $\mathcal{O}\left((k/\calH)^2\right)$ and hence vanish in
the superhorizon limit.

\section{formulas for arbitrary matter independent of gravity}

Now, let us derive a few useful formulas valid for any gravity theory.
Independent of the theory of gra\-vi\-ty, the energy-momentum conservation must
hold, which follows from the matter equations of motion and general covariance. 

We set the perturbed metric as
\begin{eqnarray}
ds^2&=&a^2\Bigl[-(1+2A)d\eta^2+2\partial_jB dx^jd\eta\nn\\
&& \qquad
+\left\{\delta_{ij}(1+2\calR)+2\partial_i\partial_j E\}dx^idx^j\right\}
\Bigr]\,,
\label{metric}
\end{eqnarray}
and the perturbed energy-momentum tensor as
\begin{align}
T^0_0&=-(\rho+\delta\rho)\,,
\quad
T^0_j=(\rho+P)u^0u_j=\frac{\rho+P}{a}u_j\,,
\nonumber\\
\quad
T^i_j&=(P+\delta P)\delta^i_j+\Pi^i{}_j\,;\quad \Pi^k{}_k\equiv0\,.
\end{align}
For a scalar-type perturbation, $u_j$ can be written as
a spatial gradient,
\begin{align}
u_j=-a\partial_j(v-B)\, \qquad \rightarrow \qquad T^0_j=-(\rho+P)\partial_j(v-B) \,
\end{align}
$\Pi^k{}_j$ in the form can be written as
\begin{align}
\Pi_{ij}=\delta_{ik}\Pi^k{}_j
=\left[\partial_i\partial_j-\frac{1}{3}\delta_{ij}\lapl\right]\Pi\,,
\end{align}
where $\lapl=\delta^{ij}\partial_i\partial_j$.

In this work, we mainly consider the following gauge-invariant variables:
\begin{align}
\calR_c &\equiv \calR - \calH(v-B)\,,
\\
\zeta &\equiv \calR -\frac{\calH}{\rho'} \delta \rho
=\calR+\frac{\delta\rho}{3(\rho+P)}\, ,
\\
V_f &\equiv (v-B)-\frac{\calR}{\calH}\,.
\end{align}
Their geometrical meanings are apparent: $\calR_c$ represents
the curvature perturbation on comoving slices ($v-B=0$), $\zeta$ the
curvature perturbation on uniform density slices ($\delta \rho =0$), and
$V_f$ the velocity potential on flat slices ($\R=0$).
They are related to each other as
\begin{align}
&\calR_c=-\calH V_f\,,
\label{calRcVf}\\
&\zeta\equiv\calR_{ud}=\calR_c+\frac{\delta\rho_c}{3(\rho+P)}\,.
\label{zetacalRc}
\end{align}
There relations will become useful later. Hereafter we use the suffix `$c$'
for quantities on comoving slices, the suffix `$ud$' for those on
uniform density slices, and the suffix `$f$' for those on flat slices.

The equation of motion is given by $\delta(\nabla_\mu T^\mu{}_j)=0$.
Explicitly we have
\begin{align}
(\rho+P)\bigl[\partial_j(v-B)'
+&\calH(1-3c_w^2)\partial_j(v-B)\bigr]
\nonumber\\
=(\rho+P)\partial_jA&+\partial_k(\delta T^k{}_j)
\nonumber
\\
\qquad\qquad\qquad
=(\rho+P)\partial_jA&+\partial_j\delta P+\frac{2}{3}\partial_j\nabla^2\Pi\,.
\label{eom}
\end{align}
Therefore, we may remove the common partial derivative $\partial_j$
to obtain
\begin{align}
&&(\rho+P)\left[(v-B)'+\calH(1-3c_w^2)(v-B)\right] \nn\\
&&\qquad=(\rho+P)A+\delta P+\frac{2}{3}\nabla^2\Pi.
\end{align}

On comoving slices, $v-B=0$ ($\Leftrightarrow~T^0{}_j=0$).
Hence
\begin{align}
(\rho+P)A_c+\delta P_c+\frac{2}{3}\nabla^2 \Pi=0\,.
\end{align}
If the perturbation is adiabatic, 
by definition $\Pi=0$. 
Thus we find
\begin{align}
\delta P_c=-(\rho+P)A_c\,.
\label{Pceq}
\end{align}
Note that this relation between $\delta P_c$ and $A_c$
is completely independent of the theory of gravity.

\section{Useful relations among gauge-invariant variables independent of gravity}

Combining Eqs.~(\ref{cssq}), (\ref{dpnad}) and (\ref{Pceq}),
we now have
\begin{align}
\delta P_{nad}=(c_s^2-c_w^2)\delta\rho_c=\frac{c_w^2-c_s^2}{c_s^2}(\rho+P)A_c\,.
\label{PnadAc}
\end{align}
The first equality is an identity, while the second comes from the 
conservation of the energy momentum tensor, and is valid for any gravity theory.
This equation may be regarded as a statement that $\delta P_{nad}$
has the same behavior as $\delta\rho_c$ and $A_c$ unless $c_w^2=c_s^2$.
In other words, the proper-time slicing ($A=0$), comoving slicing ($v-B=0$) and 
uniform density slicing ($\delta\rho=0$) coincide with each other (approximately) 
if $c_w^2\neq c_s^2$ and $\delta P_{nad}=0$ (approximately). Namely,
\be
\{\delta\P_{nad}\approx 0, c_s\neq c_w \}
 \Rightarrow \delta\rho_c\approx A_c\approx 0\,.
\ee

We can use Eq.~(\ref{PnadAc}) to obtain for example a general relation 
between the comoving curvature perturbation $\R_c$ and uniform density 
curvature perturbation $\zeta$,
\be
\zeta=\R_c -\frac{H}{\dot{\rho}}\delta\rho_c
=\R_c+\delta P_{nad}\frac{H}{\dot{\rho}(c^2_w-c^2_s)} 
\label{zetaR}
\ee
This is in agreement with the well-known coincidence of $\zeta$ and $\R_c$ 
on super-horizon scales for slow roll-models in general relativity, 
since in this case $c_s\neq c_w$ and $\delta P_{nad}\approx 0$ on superhorizon
scales. Note also that this relation is degenerate in the case of $c_s=c_w$.
As an example of such a case during inflation, later we explicitly 
consider 
the so-called ultra-slow roll inflation model.

\section{Formulas for arbitrary matter in general relativity}
Here we focus on the case of general relativity.
On comoving slices, the $G^0_0$- and $G^0_i$-components of the perturbed 
Einstein equations give  
\begin{align}
\lapl\left[\calH\sigma_c+\calR_c\right]=-4\pi G\delta\rho_c\,,
\label{G00}
\\
\calR_c'=\calH A_c\,,
\label{G0i}
\end{align}
where $\sigma$ denotes the scalar shear: $\sigma \equiv B-E'$. 
The $G^i_j$-components give, for adiabatic perturbations $\Pi=0$
and $\delta P_c=c_s^2\delta\rho_c$,
\ba
\frac{2}{a^2}(\calH'-\calH^2)A_c&=&8\pi Gc_s^2\delta\rho_c\\
\sigma_c'+2\calH\sigma_c+A_c+\calR_c&=&0.
\ea
Using the Friedman equation we then derive the equation of motion for $\R_c$:
\begin{align}
\calR_c''+\frac{{z^2}'}{z^2}\calR_c'-c_s^2 \lapl\calR_c=0\,;
\quad z^2\equiv 
\frac{(\rho+P)a^4}{c_s^2\calH^2}\,.
\label{calRceq}
\end{align}

Substituting Eq.~(\ref{G0i}) in  Eqs.~(\ref{PnadAc}) and (\ref{zetaR}) now gives 
\ba
\delta P_{nad}&=&
 \left[\left(\frac{c_w}{c_s}\right)^2-1\right](\rho+P)\frac{\dot{\R_c}}{H}
\label{PnaddotRc}
 \\
\zeta&=&\R_c-\frac{\dot{\R_c}}{3c_s^2H}.
\label{equival}
\ea
Thus $\delta P_{nad}=0$ if either $c_w^2=c_s^2$ or $\dot\calR_c=0$.
In particular in the latter case, $\dot\calR_c=0$, we have
$\zeta=\calR_c$.

\subsection{Conserved $\zeta$ and adiabaticity}
Here we briefly review the common notion \cite{Malik3} that the 
superhorizon conservation of $\zeta$ follows directly from adiabaticity, 
independent of gravity. Indeed, demanding $\delta(\nabla_\mu T^\mu_0)=0 $ yields,
 in the uniform density slicing,
\ba
\zeta'=-\frac{\calH\delta P_{nad}}{(\rho+P)}+\frac{1}{3}\lapl\left(v-E'\right)_{ud}\,
\label{dzeta}
\ea

The usual interpretation of the above equation is that for 
adiabatic perturbations, $\zeta$ is conserved on super-horizon scales, 
as long as the gradient terms can be neglected. However, as we have seen,
actually adiabaticity in the general sense (as defined in Eq.~(\ref{dpnadstr})) 
does not necessarily
imply $\delta P_{nad}=0$. Furthermore, neglecting the gradient
terms may not be justified. 


In the remainder of this letter we will consider the case of a minimally
 coupled scalar field in general relativity, as an example of the 
applications of the general relations that we have just derived.

\section{Ultra slow-roll inflation}
\label{usri}
As an interesting non-trivial example in which the 
equivalence between $\R_c$ and $\zeta$ fails to hold,
we consider the ultra slow-roll inflation (USR): a minimally coupled single 
scalar field model with constant potential. 

When $V=V_0$, the background 
scalar field equation becomes $\ddot{\phi}+3H\dot\phi=0$, and
the density and pressure perturbations become equal to each other,
$\delta P=\delta\rho$, in arbitrary gauge. Therefore we have
\begin{align}
c_w^2=c_s^2=1 \,,\quad
\delta P_{nad}=\delta P_{c,nad}=0\,.
\end{align}
In other words, the perturbation is adiabatic both in the sense of
$\delta P_{nad}=0$ and $\delta P_{c,nad}=0$.
Solving the background equations, we obtain
\be
\dot{\phi} \propto a^{-3}\,.
\ee
In particular, this implies $H=const.$ is an extremely good approximation
except possibly for the very beginning of the ultra slow-roll phase.
This gives
\be
\eps \equiv -\frac{\dot{H}}{H^2} = \frac{\dot{\phi}^2}{2H^2} \propto a^{-6}, 
\qquad \delta \equiv \frac{\ddot\phi}{H\dot\phi}
=\frac{1}{2}\frac{\dot{\eps}}{\eps H} = -3\,.
\ee

We are now in the position to appreciate the peculiarity of ultra slow-roll 
inflation. 
Let us reconsider the relations we found in the previous section. 

First, as we saw in Eq.~(\ref{PnaddotRc}) $\delta P_{nad}=0$
implies $\dot\calR_c=0$ if $c_s^2\neq c_w^2$. However, since we have
$c_s^2=c_w^2=1$ in ultra slow-roll inflation, we are unable to claim
anything about the conservation of $\calR_c$.

Second, the comoving slicing coincides with the uniform density 
slicing (and $\R_c$ with $\zeta$)
if $\dot\calR_c=0$, see Eq.~(\ref{equival}). However, again, we are unable
to claim anything since we do not know if $\calR_c$ is conserved or not.
In fact, we find that $\calR_c$ is not conserved even on superhorizon scales. 
The same follows from Eq.~(\ref{zetaR}): when $c_s^2=c_w^2$ that relation is 
degenerate, so $\zeta$ and $\R_c$ do not necessarily coincide.

Third, we concluded from Eq. \ref{dzeta} that
$\zeta$ is conserved
on superhorizon scales if $\delta P_{nad}=0$. However, as noted there,
this is true only if the gradient terms are negligible. As we shall
see below it happens that here they are not negligible at all.

\subsection{$\zeta$ and $\calR_c$ in ultra slow-roll inflation}
\label{compzeta}

From Eq.~(\ref{equival}), we have 
\begin{align}
\zeta=\calR_c-\frac{\dot\calR_c}{3H}
=-\frac{a^3}{3H}\partial_t \left(\frac{\calR_c}{a^3}\right)\,.
\label{zetausl}
\end{align}
From Eq.~(\ref{calRceq}), on superhorizon scales,
we find that the time derivative of the time-dependent solution is given by
\begin{align}
\dot\calR_c\propto\frac{1}{az^2}=\frac{H^2}{\dot\phi^2a^3}\propto a^3\,.
\end{align}
Since $H$ is almost constant in USR, we conclude that $\calR_c$ is
not conserved but grows as $a^3$ on superhorizon scales. Inserting this
to Eq.~(\ref{zetausl}) implies $\zeta=0$. Thus it seems that $\zeta$ is 
still conserved (corresponding to the conserved solution of $\calR_c$)
and the rapidly growing solution of $\calR_c$ does not contribute to $\zeta$
at all.

The above conclusion, however, is valid only in the strict large scale limit.
The finiteness of the wavelength can affect the behavior of the perturbation 
significantly even if the wavelength is much larger than the Hubble horizon
size. To see this, one can take into account the spatial gradient term of
 Eq.~(\ref{calRceq}) iteratively. For simplicity, we work in the Fourier space
where we replace $\lapl$ by $-k^2$. The superhorizon solution for $\R_c$ is then
\be
\R_c=c_1\left(1+\mathcal{O}(k^2)\right)
+c_2 a^3\left(1+\frac{1}{2}\frac{k^2}{\calH^2}+\mathcal{O}(k^4)\right)\,.  
\label{rsh}
\ee
Inserting this into Eq.~(\ref{equival}) gives
\be
\zeta = c_1\left(1+\mathcal{O}(k^2)\right) 
+\frac{c_2a^3}{3}\left(\frac{k^2}{\calH^2}+\mathcal{O}(k^4)\right).
\label{zetasol2}
\ee
Thus we see that the time-dependent solution grows like $a$ even 
on superhorizon scales. More specifically, $\zeta(t)\approx \zeta(t_k) a(t)/a(t_k)$
where $t_k$ is the horizon crossing time $a(t_k)=kH$ of the wavenumber $k$.

\begin{table*}
  \begin{tabular}{|l|l|l|}
  \hline
  & {Any Gravity theory } & {General Relativity ($A_c=\dot{\R_c}/H$)} \\ 
  \hline
 Generic matter & $\delta P_{\rm nad}
 =\delta \rho_c(c_s^2-c_w^2)
 =\left[\left(\frac{c_w}{c_s}\right)^2-1\right] (\rho+P) A_c
 $ &  
$\delta P_{\rm nad}= \left[\left(\frac{c_w}{c_s}\right)^2-1\right](\rho+P)\frac{\dot{\R_c}}{H} $   
  \\ \hline 
   M. c. scalar field & $\delta P_{\rm nad}=(c^2_w-1)A_{\rm c} \dot{\phi} ^2$   & $\delta P_{\rm nad}=(c^2_w -1 )\frac{ \dot{\R_c} }{H}\dot{\phi}^2$ 
  \\ \hline 
  \end{tabular}
\newline
\vspace*{1 cm}
\newline
\centering
  \begin{tabular}{|l|l|l|}
  \hline
  & {Any Gravity theory } & {General Relativity } \\ 
  \hline
 Generic matter & $\zeta=\R_c-\delta P_{\rm nad}\frac{H}{\dot{\rho}(c^2_s-c^2_w)} = \R_c +\frac{H}{\dot{\rho}} \frac{\rho+P}{c_s^2} A_c$ & $  \zeta=\R_c+(\rho+P)\frac{\dot{\R_c}}{c_s^2H}\frac{H}{\dot{\rho}} $
  \\ \hline 
   M. c. scalar field & $ \zeta=\R_c+ A_{\rm c} \dot{\phi} ^2\frac{H}{\dot{\rho}} \,   $& $\zeta=\R_c+\dot{\phi}^2\frac{\dot{\R_c}}{H}\frac{H}{\dot{\rho}} $  
  \\ \hline 
  \end{tabular}
  \caption{The upper table shows the relation between the fluid-based 
non-adiabatic pressure perturbations $\delta P_{nad}$ and metric perturbations, 
and the lower table gives the relation between curvature perturbations on uniform 
density slices $\zeta$ and on comoving slices $\R_c$. 
For both tables the first column corresponds to relations valid in any 
gravity theory, the second column to the case of general relativity, 
the first row is for a generic matter field and the second one is for a 
minimally coupled scalar field.}
 \end{table*}

\section{Discussion and conclusions}

The seminal works \cite{Malik3,lms} have taught us that for any 
relativistic theory of gravity, adiabaticity implies that $\zeta$ and $\R_c$ 
coincide and are conserved when gradient terms can be neglected, which in 
general happens on superhorizon scales. In this work, we have provided more
 insight into this claim.

First, we have specified that the above statement holds when (non)-adiabaticity
 is defined in the thermodynamical sense, see Eq.~(\ref{dpnadther}). 
We have argued that for a system out of equilibrium, like the early universe, 
one should define (non)-adiabaticity in the strict sense, as in Eq.~(\ref{dpnadstr}). 
In this work, we have looked at perturbations which are strictly adiabatic in 
that strict sense ($\delta P_{c,nad}=0$), and checked the implications for 
non-adiabaticity in the thermodynamical sense $\delta P_{nad}$. A third 
definition of non-adiabaticity states that the adiabatic limit has been reached
 as soon as the time-dependent solution (the non-freezing one) for $\zeta$ has
 become totally negligible.

Second, we have rewritten the relation between (thermodynamical) non-adiabaticity 
and conserved quantities in such a way as to clarify when exactly gradient terms can
 be neglected, bypassing the need for an explicit computation of these gradient terms.
 In Eq.~(\ref{PnadAc}) we have shown that for any gravity theory, $\delta P_{nad}$ 
is proportional to the lapse function in comoving slicing, $A_c$, provided that 
$c_s^2\neq c_w^2$. In the particular case of general relativity, 
$A_c$ is proportional 
to $\dot{\R}_c$ so we obtain the proportionality between $\delta P_{nad}$ and 
$\dot{\R}_c$, still under the condition that $c_s^2\neq c_w^2$. Furthermore, 
we have obtained in Eq.~(\ref{zetaR}) that when $\delta P_{nad}=0$, $\R_c$ and 
$\zeta$ coincide, again under the condition that  $c_s^2\neq c_w^2$. This results 
holds independently of gravity theory as well.

As an illustration, finally, we have studied the model of ultra slow-roll (USR) 
inflation, where $\delta P_{c,nad}=\delta P_{nad}=0$ and $c_w=c_s=1$. Indeed, 
for USR inflation all relations above obtained break down: $\zeta$ and $\R_c$ 
do not coincide and are both not conserved.
This is an example of the fact that adiabaticity (in the thermodynamic sense) is
 not always enough to ensure 
the conservation of $\R_c$ or $\zeta$.\\
\\

\noindent\paragraph{\bf\emph{Acknowledgments}} 
It is a pleasure thank Karim Malik, Jorge Nore\~na, Gonzalo Palma, 
Sergey Sibiryakov and Drian van der Woude for illuminating discussions. 
This work was supported by the Fondecyt 2015 Postdoctoral Grant 3150126 (SM),
 the ``Anillo'' project ACT1122 funded by the ``Programa de Investigaci\'on 
Asociativa" (SM), 
and by the the  Greek national funds under the  ``ARISTEIA'' Action, 
the Dedicacion exclusica and Sostenibilidad programs at UDEA, 
the UDEA CODI projects IN10219CE and 2015-4044
(AER), and in part by MEXT KAKENHI Grant Number 15H05888.

\end{document}